\documentclass[preprint,aps,nofootinbib,preprintnumbers,amsmath,amssymb,11pt]{revtex4}
\usepackage{epsfig}
\graphicspath{ {images/} }
\usepackage{amsmath}
\usepackage{mathtools}
 \usepackage{multirow}
\usepackage{xcolor}
\usepackage{slashed}
\usepackage{simplewick}
\usepackage{amssymb}
\usepackage{hyperref}  
\usepackage{amsfonts}
\usepackage{graphicx}
\usepackage{dcolumn}
\usepackage{bm}
\usepackage{natbib}

\newcommand{\be}{\begin{equation}}
\newcommand{\ee}{\end{equation}}
\newcommand{\bea}{\begin{eqnarray}}
\newcommand{\eea}{\end{eqnarray}}

\newcommand{\nn}{\nonumber}

\def\cN {{\cal N}}

\def\e{\epsilon}

\begin{document}


\title{Bootstrapping the $\mathcal{N}=1$ Wess-Zumino models in three dimensions}
\author{\vspace*{1 cm}\large
Junchen Rong$^{\,\clubsuit}$, Ning Su$^\dagger$}
\affiliation{\vspace*{0.5 cm}$^\clubsuit$ DESY Hamburg, Theory Group, \\Notkestraße 85, D-22607 Hamburg, Germany,\\}
\affiliation{$^\dagger$ 
Institute of Physics, \'Ecole Polytechnique F\'ed\'erale de Lausanne, CH-1015 Lausanne, Switzerland\\\\\\}

\begin{abstract}
\vspace*{1 cm}
Using numerical bootstrap method, we determine the critical exponents of the minimal three-dimensional $\mathcal{N}=1$ Wess-Zumino models with cubic superpotetential $\mathcal{W}\sim d_{ijk}\Phi^i\Phi^j\Phi^k$. The tensor $d_{ijk}$ is taken to be the invariant tensor of either permutation group $S_N$, special unitary group $SU(N)$, or a series of groups called $F_4$ family of Lie groups. Due to the equation of motion, at the Wess-Zumino fixed point, the operator $d_{ijk}\Phi^j\Phi^k$  is a (super)descendant of $\Phi^i$. We observe such super-multiplet recombination in numerical bootstrap, which allows us to determine the scaling dimension of the super-field $\Delta_{\Phi}$.
\end{abstract}

\vspace*{3 cm}
\maketitle

\def\thesection{\arabic{section}}
\def\thesubsection{\arabic{section}.\arabic{subsection}}
\numberwithin{equation}{section}
\newpage

\section{Introduction}
\label{Introduction}
Higher dimensional bootstrap program is a generalization of the extremely successful two dimensional conformal bootstrap program \cite{Polyakov:1974gs,Ferrara:1973yt,Mack:1975jr}. Initialized by the seminal work \cite{Rattazzi:2008pe},  it has become an important method to study conformal field theories in various space time dimensions, see \cite{Poland:2018epd} for a recent review. 

Recently, it was realized \cite{Rong:2018okz,Atanasov:2018kqw} that the minimal superconformal field theory in three space-time dimension could be studied using the bootstrap method. The Lagrangian of the theory is given by
\be\label{Lagan}
\mathcal{L}=\frac{1}{2}(\partial_{\mu} \sigma)^2+\bar{\psi} \slashed{\partial}\psi+\frac{\lambda}{2}\sigma\bar{\psi}{\psi}+\frac{\lambda^2}{8}\sigma^4,
\ee
with $\sigma$ and $\psi$ being a scalar and a Majorana spinor respectively. The theory is invariant under time reversal symmetry (T-parity)  under which $\sigma\rightarrow -\sigma$ and $\psi \rightarrow \gamma^0\psi$. The Lagrangian can be written as a Wess-Zumino model with superpotential $W=\lambda \Sigma^3$. Here $\Sigma=\sigma+\bar{\theta}\psi+\frac{1}{2}\bar{\theta}\theta \e$ is a real superfield.  
To bootstrap this theory, one need to consider four point correctors $\langle \sigma \sigma\sigma\sigma\rangle$,  $\langle \epsilon \epsilon\epsilon\epsilon\rangle$ and $\langle\sigma \sigma \epsilon \epsilon\rangle $, with $\epsilon$ being the superconformal descendant of $\sigma$ ($\epsilon$ is a conformal primary). The operator product expansion (OPE) coefficients in $\sigma \times \sigma$, $\sigma \times \epsilon$ and $\epsilon \times \epsilon$ are related to each other, and the relation is fixed by superysmmetry. This is a generalization of the ``long multiplet bootstrap'' idea \cite{Cornagliotto:2017dup} used in two dimensional superconformal bootstrap.  By assuming the fact that the theory has only one T-parity even relevant scalar conformal primary,  $\Delta_{\sigma}$ was determined to high precision.  A bootstrap island was discovered if one further assume that the theory has only two T-parity odd relevant scalar conformal primaries.  Notice unlike SCFTs with higher number of supersymmetry, $\cal N$=1 SCFT has no R symmetry, and the scaling dimension of $\Phi$ can not be determined exactly by analytic methods.

In the present paper, we generalize our method to study $\cN=1$ superconformal field theories (SCFTs) with global symmetries. To be more specific, we determine the scaling dimension of the superfield $\Phi^i$ in various types of Wess-Zumino models with a cubic super-potential 
\be
\mathcal{W}\sim d_{ijk}\Phi^i\Phi^j\Phi^k.
\ee
Here we choose $d_{ijk}$ to be the invariant tensor of certain flavor symmetry group.  As a result of equation of motion, the operator $d_{ijk}\Phi^j\Phi^k$  would be a (super)descendants of $\Phi^i$. This is the most essential property that distinguishes the Wess-Zumino models from other CFT's. For example, in generalized free theories, $d_{ijk}\Phi^j\Phi^k$  is a superconformal primary. Such a phenomena should be viewed as a supersymmetric version of the multiplet recombination appeared in 3d Ising CFT, where $\sigma^3$ recombines with $\sigma$ at the fixed point \cite{Rychkov:2015naa}.  

\section{general discussion}
Conformal multiplets in $\cal N$=1 superconformal field  theories group themselves into super-multiplets.  There are in total four types of multiplets, which we denote as $\mathcal{B}^{l}_{+}$, $\mathcal{B}^{l}_{-}$, $\mathcal{F}^{j}_{+}$ and $\mathcal{F}^{j}_{-}$ as in \cite{Rong:2018okz}. ``$\mathcal{B}/\mathcal{F}$'' tells us whether the super-primary field is bosonic or fermionic. A generic super-multiplet contains four conformal multiplets, suppose the superconformal primary has spin $l$ and scaling dimension $\Delta_0$, there are two level-1 (super)descendant with $\Delta=\Delta_0+1/2$ and spin $l\pm1$. There is also a level-2 descendant with $\Delta=\Delta_0+1$ and spin $l$.

For bosonic super-fields $\mathcal{B}^{l}_{\pm}$, the subscript ``$+/-$'' denotes the parity of the super-primary\footnote{More precisely speaking, what we call parity here really should be called ``twised parity'', which is defined as $(-1)^l \times (\text{parity})$.  We call operators of the form $\phi\partial^{\mu_1}\ldots \partial^{\mu_l}\phi$ to be parity even (denoted as ``$+$''), and operators of the form $\phi\partial^{\mu_1}\ldots \partial^{\mu_l}F$ to be parity odd (denoted as ``$-$''), suppose $\phi/F$ is parity odd/even. This convention is convenient for us as it tells us whether an operator could appear in the OPE of two parity even scalar operators.}.  $l$ denotes the spin of the super-primary. The superconformal primary and the level-2 descendant have opposite parity. In this notation, the superfield $\Phi$ that appears in the Lagrangian is a $\mathcal{B}^{0}_{-}$ multiplet.

For fermonic super-fields $\mathcal{F}^{j}_{\pm}$, remember the multiplet contains two level-1 super-descendant (but conformal primary) fields, with spin $j\pm1$. The subscript ``$+/-$'' denotes the parity of the level-1 descendant with spin $j-\frac{1}{2}$. The operator with spin $j+1$ has opposite parity.

Superconformal primary operators should also carry flavor symmetry indices and transform in certain irreps of the flavor symmetry group. As mentioned in the introduction, the SCFT we consider will be the infra-red fixed point of the free $\cal N$=1 theory deformed by the super-potential $\mathcal{W}\sim d_{ijk}\Phi^i\Phi^j\Phi^k$. The flavor symmetry group therefore need to admit such an fully symmetric invariant tensor $d_{ijk}.$ We will consider three types of flavor symmetry groups that satisfies this condition, which are the permutation groups $S_{n+1}$, the special unitary groups $SU(N)$, and the so  called $F_4$ family of Lie groups.  We use the label ``n'' to denote the irrep that $\Phi^i$ transforms in, the precise meaning of ``n'' depends on the flavor symmetry group considered and is summarized in Table \ref{ntimesn}.
\begin{table}[h]
\begin{tabular}{|l|c|c|}\hline\hline
sym. group                & dim. of n & Irrep name\\\hline
$S_N$  & $N-1$  & fundamental  \\ \hline
$SU(N)$ & $N^2-1$   &  adjoint \\ \hline
$F_4$-family & Table \ref{F4table}   &  Table \ref{F4table}   \\ \hline
\end{tabular}
\caption{Flavor symmetry groups that we consider.}\label{ntimesn}
\end{table}

\begin{figure}[h]
\includegraphics[width=12cm]{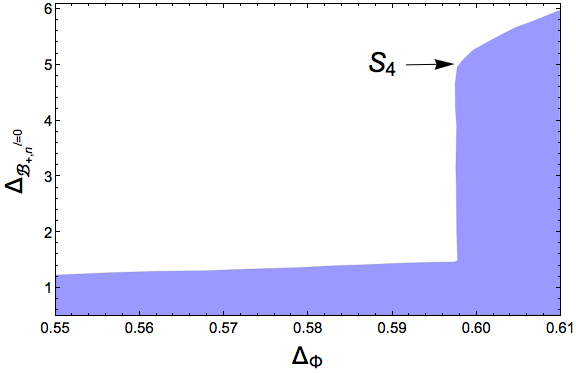}
\caption{Bound on the scaling dimension of the leading operator transforming in $3$-dimension of $S_4$, which is T-parity even. The numerics is preformed at $\Lambda=13$.}\label{cliff}
\end{figure}

Notice as a result of the invariant tensor $d_{ijk}$, $n\otimes n$ contains the irrep $n$ itself:
\be
n\times n \rightarrow S+ n +\ldots.
\ee
The ``$\ldots$'' part depends on which family of flavor symmetry group you are considering.  Taking into account also the superconformal symmetry, we have 
\be
\mathcal{B}^{l=0}_{-,n}\times \mathcal{B}^{l=0}_{-,n} \rightarrow \mathcal{B}^{l=0}_{+,S}+ \mathcal{B}^{l=0}_{+,n}  +\ldots.
\ee
The ``$\dots$'' contains other operators with different flavor symmetry representation and spin. We will leave the details of the bootstrap equations to Appendix \ref{nonSUSYeqn} and \ref{susyeqn}, and describe here only the numerical results. In (generalized) free theory, the leading $\mathcal{B}^{l=0}_{+,n} $ is simply the operator $d_{ijk}\Phi^j\Phi^k$. Using conformal bootstrap to bound the leading $\mathcal{B}^{l=0}_{+,n} $ in the theory studied, typically one get a plot like Figure \ref{cliff}.
As we increase $\Delta_{\Phi}$ from the unitarity bound, before reaching the Wess-Zumino model, the bootstrap is very close to the line $\Delta_{\mathcal{B}^{l=0}_{+,n}}\approx 2\Delta_{\Phi}$, resembling the fact that the spectrum of the theories saturating the bound are not too much different from the generalized free theory. As soon as $\Delta_{\Phi}$ approaches the Wess-Zumino point,  $d_{ijk}\Phi^j\Phi^k$ becomes a descendant of $\Phi^i$. Since the bootstrap bound only constrains the scaling dimension of superconformal conformal primaries, we get a much higher bound, which corresponding to the dimension of the $\mathcal{B}^{l=0}_{+,n}$ operator next to  $d_{ijk}\Phi^j\Phi^k$ . Such jumps was observed in all the flavor symmetry groups that we have considered. We identify them as the Wess-Zumino models.

Having understood that the super-multiplet recombination being the numerical signatures of the Wess-Zumino models, we focus on extracting the location of the jump of bootstrap curve. We can impose the following conditions,
\begin{itemize} 
\item $\Delta_{\mathcal{B}^{l=0}_{+,n}}\geq 3$,
\end{itemize} 
and preform a bisection study to determine $\Delta_{\Phi^i}$. This greatly save the the time it takes to do numerics. In practice, we found that it is also useful to impose another condition that 
\begin{itemize} 
\item The leading $\mathcal{B}^{l=0}_{-,S}$ operator has dimension bigger than 2,  
\end{itemize} 
which simply tells us that the superpotental is irrelevant at the fixed point. Since our SCFT is an infra-red fixed point of an $\cal N$=1 renormalization group flow, this is a natural condition. 

\subsection{$S_N$ bootstrap}  
\begin{figure}[h]
\includegraphics[width=12cm]{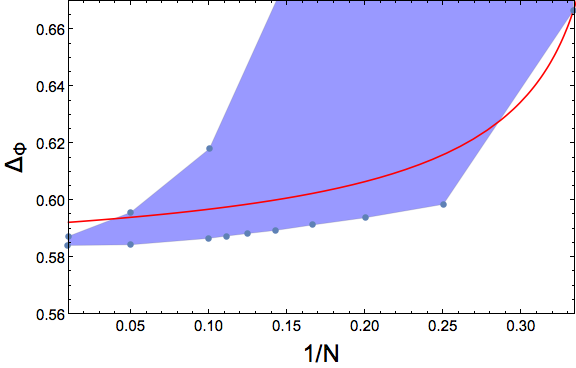}
\caption{$S_N$ invariant $\cN=1$ SCFT bootstrap. Bound on $\Delta_{\Phi}^i$ as a function of the flavor symmetry group size $N$. The red curve being the two loop prediction of $\Delta_{\Phi}$ using $\epsilon$-expansion.}\label{SNplot}
\end{figure}

Taking the flavor symmetry group to be $S_N$, and ``n'' to be the $N-1$ dimensional ``standard'' representation of $S_N$, we get Figure \ref{SNplot}.
The lower edge of the allowed region is the location of jump, hence Wess-Zumino models. 
Notice for sufficient large group size N, we obtained some ``bootstrap islands'', with $\Delta_{\Phi}$ being constrained with both upper and lower bounds. This is due to the fact that we require superpoetential to be irrelevant. Suppose we relax such a condition, the upper bound disappears. One can compare the bootstrap result with perturbative calculation using $4-\epsilon$ expansion.  The two loop $\epsilon$-expansion result for the scaling dimension of $\Phi$ is given by
\be\label{SNloop}
\Delta_{\Phi}=1+\frac{(5-3 (N-1)) \epsilon }{7 (N-1)-11}+\frac{\left(7 (N-1)^3-34 (N-1)^2+53 (N-1)-26\right) \epsilon ^2}{(7 (N-1)-11)^3}+\mathcal{O}(\epsilon^3).
\ee
The result is obtained by plugging explicit form of the invariant tensor into the beta function and anomalous dimension formulas summarised in Appendix A of \cite{Fei:2016sgs}. 
The red curve in Figure \ref{SNplot} corresponds to $\epsilon=1$. Taking into account the fact that this is only two loop result, the agreement is pretty good. Notice also as $N\rightarrow \infty$, $\Delta_{\Phi}$ approaches 0.584444(30), which is the value of $\Delta_{\Phi}$ in $\cal N$=1 Ising model \cite{Rong:2018okz}. In fact, the large N limit of \eqref{SNloop} is also given by the corresponding $\epsilon$-expansion series of $\cal N$=1 Ising model. This means that the large N limit of this $S_N$ Wess-Zumino models decouple into N copies of $\cal N$=1 Ising models. It might be possible to reach the $\cal N$=1 Potts model by deforming N copies of $\cal N$=1 Ising models with a certain relevant operator, and study $\Delta_{\Phi}$ using conformal perturbation theory and compare with the bootstrap result. See \cite{Komargodski:2016auf} for a similar story using conformal perturbation to study RG flow from non supersymmetry Ising models.

\begin{figure}[h]
\includegraphics[width=12cm]{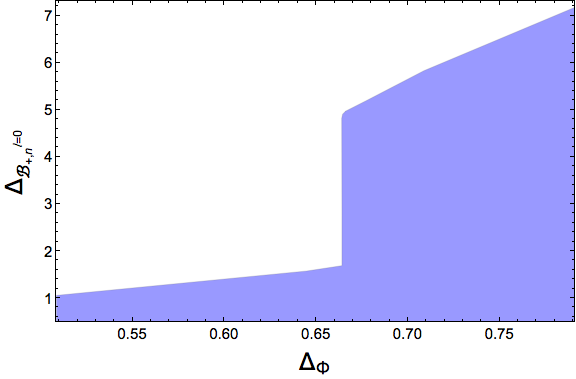}
\caption{$S_3$ invariant $\cN=1$ SCFT bootstrap. Bounds on $\Delta_{\Phi}^i$ as a function of the flavor symmetry group size $N$. }\label{S3plot}
\end{figure}

\begin{figure}[h]
\includegraphics[width=12cm]{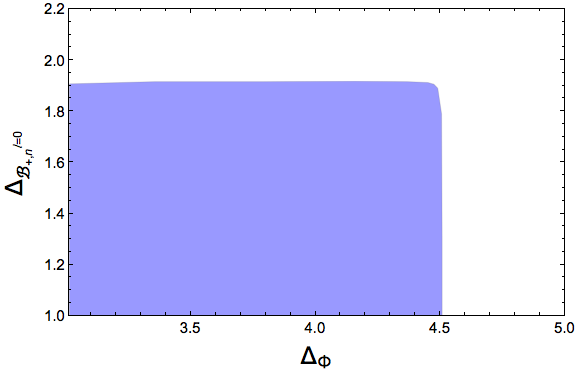}
\caption{$S_3$ invariant $\cN=1$ SCFT bootstrap. Bounds on $\mathcal{B}^{l=0}_{+,S}$ and $\mathcal{B}^{l=0}_{+,n}$ after setting $\Delta_{\Phi}=2/3$. }\label{S3plot2}
\end{figure}

The $S_3$ invariant Wess-Zumino models deserves a few more words. The  jump happens at precisely $\Delta_{\Phi}=2/3$, as shown in Figure \ref{S3plot}. This is because the $\cal N$=1 Wess-Zumino model has a supersymmetry symmetry enhancement. The group $S_3$ is a subgroup of the R-symmetry group of $\cal N$=2. When viewed as an $\cal N$=1 theory, it becomes a flavor symmetry. The corresponding SCFT is precisely the so call ``$\cal N$=2 Ising model'' studied in \cite{Bobev:2015vsa}.
Setting  $\Delta_{\Phi}=2/3$, we could study the spectrum of other operators in this SCFT, as shown in Figure \ref{S3plot2}. 
The operator $\mathcal{B}^{l=0}_{+,S}$, in $\cal N$=2 language, corresponds the leading R-charge singlet scalar operator, whose scaling dimension was shown to be 1.9098(20), using $\cal N$=2 bootstrap in \cite{Bobev:2015vsa}. The operator $\mathcal{B}^{l=0}_{+,n}$, in $\cal N$=2 language, corresponds the leading R-charge 2 scalar operator, whose dimension could also be estimated using extremal functional method \cite{ElShowk:2012hu} at the $\cal N$=2 bootstrap kink, which turns out to be around 4.38. One see explicitly that the spectrum obtained here using $\cal N$=1 bootstrap agrees with the corresponding $\cal N$=2 results. 

The above analysis of the $S_3$ bootstrap result provides further support for our identification of the jump of the bootstrap curve to the Wess-Zumino models.

\subsection{$SU(N)$ bootstrap}  

\begin{figure}[h]
\includegraphics[width=12cm]{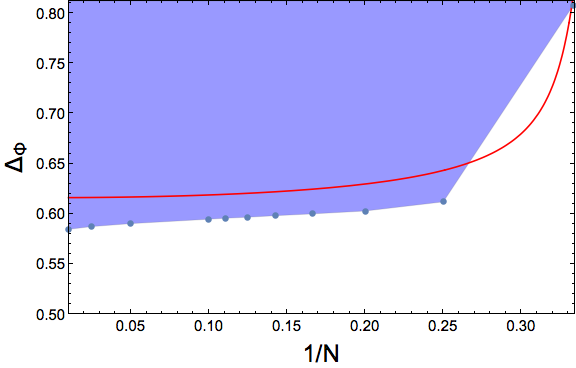}
\caption{$SU(N)$ invariant $\cN=1$ SCFT bootstrap. Bound on $\Delta_{\Phi}^i$ as a function of the flavor symmetry group size $N$. The red curve being the two loop prediction of $\Delta_{\Phi}$ using $\epsilon$-expansion.}\label{SUNplot}
\end{figure}

Taking the flavor symmetry group to be $SU(N)$, and ``n'' to be the adjoint representation of $SU(N)$, we get Figure \ref{SUNplot}. The plot has the same feature as the $S_N$ bootstrap curve, except that the irrelevance of the superpotential does not yield  a upper bound even in large N. 

The two loop $\epsilon$-expansion result for the scaling dimension of scalar is given by
\be
\Delta_{\Phi}=1-\frac{2 \left(N^2-8\right) \epsilon }{5 N^2-36}+\frac{2 \left(N^6-26 N^4+208 N^2-480\right) \epsilon ^2}{\left(5 N^2-36\right)^3}+\mathcal{O}(\epsilon^3).
\ee
The $\epsilon=1$ result again corresponds to the red curve in Figure \ref{SUNplot}. Taking into account the fact that we are this is two loop result, the agreement is again surprising good. 

To bootstrap the SU(3) theory and the $F_4$ family of theories to be discussed in the next section, there is one extra complication. We have to impose a gap in the $\mathcal{F}^{j=1/2}_{+,n}$ channel to remove the ``fake primary effect'' \cite{Karateev:2019pvw}, the details is explained in Appendix \ref{fakeprimary}. $\cal N$=1 SU(3) invariant Wess-Zumino model is of special interest since it is related to a certain ${\cal N}=1$ duality. Consider $2+1D$ $\cN=2$ QED with 2 chiral multiplets of gauge charge $+1$. The explicit symmetry of the theory is
 \be 
 SU(2)_{flav}\times U(1)_{top} \times U(1)_{R}.
 \ee
Using 3d/3d correspondence, it was pointed out in \cite{Gang:2017lsr,Gang:2018wek} that in the IR fixed point, global symmetry enhancement happens:  
 \be
 SU(2)_{flav}\times U(1)_{top}\rightarrow SU(3).
 \ee
Later, it was proposed \cite{Gaiotto:2018yjh,Benini:2018bhk}  that the theory has a dual description in terms of $\cN=1$ Wess-Zumino models with superpotential term $\mathcal{W}\sim d_{ijk} \Phi^i  \Phi^j  \Phi^k$, with $\Phi^i$ transforming in the adjoint representation of $SU(3)$.  In this description, at the IR fixed point, $\cN$=1 SUSY got enhancement into $\cN$=2 SUSY. The duality gives us a non-trivial prediction for the scaling dimension of the scalar superfield $\Phi$,
\be\label{duality}
\mathcal{N}=2 \text{ sQED} \longleftrightarrow \mathcal{N}=1 \text{ SU(3) Wess-Zumino model}.
\ee
On the QED side, $\Phi^i$ is dual to the $\cN=2$ superconformal primary in the conserved flavor current multiplet. The duality predicts that on the WZ model side $\Phi$ get strongly renormalized and become $\Delta_{\Phi}=1$.

\begin{table}[h]
\begin{tabular}{|c|c|c|c|c|c|c|c|c|c|c|}
\hline\hline
         & $\mathcal{B}^{l=0}_{S,+}$   & $\mathcal{B}^{l=0}_{Adj,+}$, 2$^{nd}$ & $\mathcal{B}^{l=0}_{t,+}$   & $\mathcal{B}^{l=1}_{a,+}$   & $\mathcal{B}^{l=1}_{Adj,+}$   & $\mathcal{B}^{l=0}_{S,-}$   & $\mathcal{B}^{l=0}_{Adj,-}$   & $\mathcal{B}^{l=0}_{t,-}$   & $\mathcal{B}^{l=1}_{a,-}$   & ,$\mathcal{B}^{l=1}_{Adj,-}$                       \\ \hline
$\Delta$ & 2.77                       & 4.88                                  & 0.566                       & 2.38                        & 3.03                          & 2.50                        & 4.28                          & 3.02                        & 3.79                        & 3.27                                               \\ \hline\hline
         & $\mathcal{F}^{j=1/2}_{S,+}$ & $\mathcal{F}^{j=1/2}_{Adj,+}$         & $\mathcal{F}^{j=1/2}_{t,+}$ & $\mathcal{F}^{j=3/2}_{a,-}$ & $\mathcal{F}^{j=3/2}_{Adj,-}$ & $\mathcal{F}^{j=3/2}_{S,-}$, 2$^{nd}$ & $\mathcal{F}^{j=3/2}_{Adj,-}$ & $\mathcal{F}^{j=3/2}_{t,+}$ & $\mathcal{F}^{j=1/2}_{a,-}$ & $\mathcal{F}^{j=1/2}_{Adj,-}$, 2$^{nd}$ \\ \hline
$\Delta$ & 5.25                        & 3.49                                  & 3.71                        & 3.52                        & 3.61                          & 6.05                        & 3.27                          & 3.05                        & 2.65                        & 5.11                                               \\ \hline
\end{tabular}
\caption{Spectrum of (sub-)leading operators in each channel, estimated at the SU(3) bootstrap jump using extremal functional method. Here``$t$'' labels the 27 dimensional representation of SU(3), whose Dynkin label is (2,2), while``$a$'' labels the 10+$\overline{10}$ dimensional representation of SU(3), whose Dynkin label is (3,0)+(0,3).}\label{spec}
\end{table}
In table \ref{spec}, we use extremal functional method to estimate the spectrum of the leading operators in each bootstrap channel. The result $\Delta_{\mathcal{B}_{+,S}}\approx 2.77$ is comparable with the corresponding two loop $\epsilon$-expansion calculation:
\be
\Delta_{\Phi^i\Phi^i}=2+0.666667 \epsilon-0.308642 \epsilon ^2+\mathcal{O}(\epsilon^3)\approx 2.358.
\ee

The fact that we observe a jump of bootstrap curve near the $\Delta_{\Phi}$ from $\epsilon$-expansion, but not the prediction from duality comes as a surprise to us. Could it be that the observed jump corresponds to another SCFT, but rather the $\cal N$=1 Wess-Zumino model given by the superpotential $\mathcal{W}\sim d_{ijk}\Phi^i\Phi^j\Phi^k$? One may consider introducing an SU(3) singlet $H$ and change the super potential to be
$$\mathcal{W}\sim \lambda_1 d_{ijk}\Phi^i\Phi^j\Phi^k+\lambda_2 H\Phi^i\Phi^i+\lambda_3 H^3.$$ This theory can not explain the bootstrap result because of the following reason. We have two $\mathcal{B}^{l=0}_{+,n}$ operators, which are $d_{ijk}\Phi^j\Phi^k$ and $H \Phi^i$. They have similar scaling dimensions, which roughly equals twice the dimension of $\Delta_{\Phi}$. After the multiplet recombination, one combination of the two operators, $\lambda_1 d_{ijk}\Phi^j\Phi^k+\lambda_2 H\Phi^i$, becomes the descendant of $\Phi^i$. We are however left with another combination of the two operators as a super-primary. This operator stops the bootstrap curve from showing a sharp jump. We should emphasis that the bootstrap result does not rule out the prediction from the proposed duality. It is possible that by studying mixed correlators bootstrap, we could settle down this issue. 

In \cite{Gaiotto:2018yjh,Benini:2018bhk}, the duality \eqref{duality} was ``derived'' using the well tested $\cal N$=4 duality between U(1) gauge theory with one hypermultiplet of charge 1 and a free massless hypermultiplet. This guarantees that anomalies match on the two sides of \eqref{duality}. It was also shown that the mass deformations lead to the same vacuum structure on the two sides. Suppose the jump of the SU(3) bootstrap curve indeed corresponds to $\cal N$=1 SU(3) invariant WZW model. To accommodate the result of \cite{Gaiotto:2018yjh,Benini:2018bhk}, one possibility is that $\mathcal{N}=2 \text{ sQED}$ and $\mathcal{N}=1 \text{ SU(3) Wess-Zumino model}$ are connected by an $\cal N$=1 time reversal symmetry preserving renormalization group (RG) flow. Since in the $\cal N$=1 Wess-Zumino theory, the leading $\mathcal{B}^{l=0}_{-,S}$ operator is irrelevant ($\Delta\approx 2.50$ as shown in Table \ref{spec}), the direction of the RG flow should be
\be
\mathcal{N}=2 \text{ sQED} \xrightarrow{\text{RG flow}} \mathcal{N}=1 \text{ SU(3) Wess-Zumino model}.
\ee
If this is indeed the case, the spectrum of $\mathcal{N}$=2 sQED, when branching into $\cal N$=1 multiplets, should contain a relevant $\mathcal{B}^{l=0}_{-,S}$ scalar. It is precisely this operator that induce the above RG flow. To check this possibility, it would be interesting to use $\cal N$=2 bootstrap to study the $\mathcal{N}$=2 sQED directly. We leave this for future work. 

\subsection{$F4$ family of Lie groups' bootstrap}

\begin{figure}[h]
\includegraphics[width=12cm]{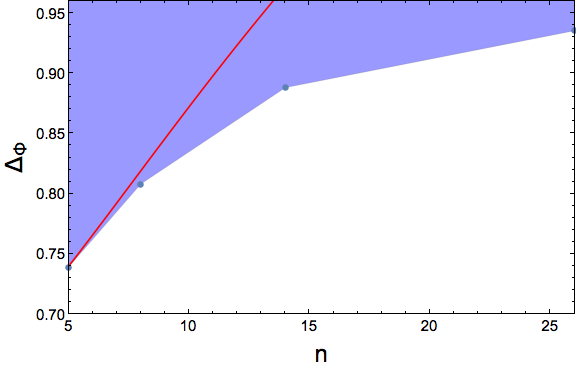}
\caption{Bootstrapping SCFT with flavor symmetry belonging to $F_4$ family. Bound on $\Delta_{\Phi}^i$ as a function of the flavor symmetry group size $N$. We have imposed the gap $\Delta_{\mathcal{F}^{j=1/2}_{+,n}}\geq \Delta_{\text{unitarity}}+1.2$ to remove the ``fake primary effect''. The red curve being the two loop prediction of $\Delta_{\Phi}$ using $\epsilon$-expansion.}\label{F4plot}
\end{figure}

The word ``$F_4$ family'' is a concept from the bird-trick classification of Lie group representations \cite{Cvitanovic:2008zz}. These group representations admits a fully symmetric invariant tensor $d_{ijk}$  satisfying the relation
\be\label{F4defination}
d_{il m}d_{mjk}+d_{ijm}d_{mkl}+d_{ikm}d_{mjl}=\frac{2\alpha}{n+2}(\delta_{ij}\delta_{kl}+\delta_{il}\delta_{jk}+\delta_{ik}\delta_{jl})
\ee
$\alpha$ is a constant that depends on the normalization of $d_{ijk}$. Such a relation allows us to write their crossing equation in a simple compact form \cite{Pang:2016xno}. The lie groups and corresponding representations that belong to this family is lised in Table \ref{F4table}.
\begin{table}[h]
\centering
\begin{tabular}{|l|c|c|c|c|}
 \hline \hline
 Group &  $F_4$ & $SO(3)$&$SU(3)$ & $Sp(6)$ \\\hline
$\Phi^I \in n$ & 26 & 5 & 8 & 14\\\hline
Irrep & fundamental  & antisymmetric traceless & adjoint & antisymmetric traceless \\
 &   & rank-2 tensor &  &  rank-2 tensor\\\hline
\end{tabular}
\caption{ $F_4$ family of invariant groups and the dimension of the corresponding of representation. The row ``Irrep'' shows the name of the irreps if a commonly used exist.}\label{F4table}
\end{table}
Taking the flavor symmetry group to a member of the $F_4$ family, and ``n'' to be the corresponding representation, we present the bootstrap result in Figure \ref{F4plot}. The $n=8$ point, corresponds to the SU(3) result that was already present in previous section, we will however leave it here for completeness. 

The two loop $\epsilon$-expansion result for the scaling dimension of scalar is given by
\bea
\Delta_{\Phi}=1-\frac{4 \epsilon }{n+10}+\frac{\left(n^3-4 n^2-4 n+16\right) \epsilon ^2}{(n+10)^3}+\mathcal{O}(\epsilon^3)
\eea
Unlike previous case for $S_N$ and $SU(N)$ bootstrap, the $\epsilon=1$ curve is in very poor agreement with the bootstrap result except for  $n=5$ and $n=8$ case. However, after checking the coefficients of the $\epsilon$-expansion series, we can see that the series shows poor convergence even at two loop order. 
\bea
\Delta_{\Phi}|_{n=5}&=&1-0.266667 \epsilon+ 0.00622222 \epsilon ^2 +\mathcal{O}(\epsilon^3),\nn\\
\Delta_{\Phi}|_{n=8}&=&1-0.222222 \epsilon+0.0411523 \epsilon ^2 +\mathcal{O}(\epsilon^3),\nn\\
\Delta_{\Phi}|_{n=14}&=&1- 0.166667 \epsilon + 0.138889 \epsilon^2+\mathcal{O}(\epsilon^3),\nn\\
\Delta_{\Phi}|_{n=26}&=&1 - 0.111111 \epsilon + 0.316872 \epsilon^2+\mathcal{O}(\epsilon^3).
\eea
We have also checked that naive Pad\'e approximation does not improve the agreement. We expect higher loop result with proper re-summation will improve the agreement with the bootstrap result. 

\section{Discussion} 
We have shown that by studying the supermultiplet recombination phenomena associated $\cal N$=1 Wess-Zumino model with a superpotential $\mathcal{W}\sim d_{ijk}\Phi^i\Phi^j\Phi^k$, we were able to determine the scaling dimension of the superfield  $\Phi$ numerically.
One important motivation to bootstrap $\cal N$=1 superconformal field theories is to study the various $\cal N$=1 dualities proposed in \cite{Benini:2018umh,Gaiotto:2018yjh,Benini:2018bhk}. It is possible that such a supermultiplet recombination phenomena will help us identify some of the Wess-Zumino models. Let us mention one example here. It was proposed in \cite{Benini:2018umh} that there exist a duality between the following two theories \newline
$$
\text{\begin{tabular}{ccc}
$U(1)_{\frac{1}{2}}$ with one flavor $Q$    & \multirow{2}{*}{$\longrightarrow$} & Wess-Zumino model with $P$ and $H$         \\
$\mathcal{W}\sim Q^{\dagger}QQ^{\dagger}Q$ &                                & $\mathcal{W}=\lambda_1 HP^{\dagger}P+ \lambda_2 H^3$
\end{tabular}
} $$
The LHS is a super QED with one flavor of Q, while the right hands is a Wess-Zumino model one real and on complex scalar superfields. The RHS has an explicit U(1) flavor symmetry, which on the LHS corresponds to the topological U(1) charge that comes from the Bianchi identity of the U(1) gauge field.  Notice the RHS theory has time-reversal symmetry which is explicitly broken on the LHS. For the duality to work, as emphasized in \cite{Benini:2018umh}, the infra-red fixed point needs to have emergent time-reversal symmetry. Or equivalently, one of the two combinations of $H^2$ and $P^{\dagger} P$ needs to be irrelevant (the other combination becomes descendant of $H$), and hence has scaling dimension bigger that 2. The two loop result studied in \cite{Benini:2018umh} gives us a pretty dangerous value, $\Delta\approx 2.038$. It would be very interesting to check this number non-perturbatively using numerical bootstrap. From the superpotential, we see that the multiplet recombination that happens in the SCFT is $HP$ combines with $P$.  We therefore need to consider the mixed correlators bootstrap of $\langle HHHH \rangle$, $\langle P^{\dagger} P P^{\dagger} P \rangle$ and $\langle P^{\dagger} PHH \rangle$. We hope to study this problem in future.

Another interesting generalization of this work is to bootstrap $\cal N$=1 SCFT's with broken time reversal symmetry. These include interesting theories such as various $\cal N$=1 Chern-Simons matter theories. 

Recently, there has been a revived interest in searching for new fixed points of scalar field theory with quartic interactions \cite{Osborn:2017ucf,Rychkov:2018vya}. It would be interesting to generalize them and search for new $\mathcal{N}=1$ fixed points. This would involve searching for flavor symmetry groups whose representations admit a symmetric invariant tensor, and analysis the 1-loop beta function to search for unitary fixed point. We leave this for future work.

\vskip 0.4 in
{\noindent\large  \bf Acknowledgments}
\vskip 0.in
We are grateful to Jin-Beom Bae, Jeong-Hyuck Park, Andy Stergiou, David Simmons-Duffin, Slava Rychkov, Zohar Komargodski, David Poland and Alessandro Vichi for helpful discussion or comments. J. R. would like to thank the hospitality of Asia Pacific Center for Theoretical Physics while the early stage of the work was finished. N. S. is supported by the European Research Council Starting Grant under grant no. 758903 and Swiss National Science Foundation grant no. PP00P2-163670. To calculate the conformal block functions, we used the code from JuliBootS program\cite{Paulos:2014vya}. The numerics is solved using SDPB program\cite{Simmons-Duffin:2015qma}. Some of the computations in this paper were run on the HPC Cluster of Institute of Theoretical Physics, Chinese Academy of Sciences.

\appendix 
\section{Fake primary effect}\label{fakeprimary}
In \cite{Karateev:2019pvw}, it was discovered that when we do non supersymmetric bootstrap, the numerical generated conformal bootstrap for internal $l=1$ operators, when $\Delta^{l=1}\rightarrow \Delta^{l=1}_{\text{unitarity}}$, looks like the conformal block for a conserved spin-1 current plus a scalar block with scaling dimension $\Delta=\Delta^{l=1}_{\text{unitarity}}+1$. This effectively allows a scalar operator with $\Delta^{l=0}=\Delta^{l=1}_{\text{unitarity}}+1$ to appear in the numerical spectrum.
This ``fake primary effect'' usually cause the bootstrap curve to show jump similar to those observed in our $\cal N$=1 bootstrap program. The reason being the following: in D=3, suppose one want to bound the leading scalar operator in certain flavor symmetry channel ``x'', one typically get a bootstrap curve in the $\Delta_{\rm x}$-$\Delta_{\rm external}$ plane. As long as the curve is below $\Delta^{l=1}_{\text{unitarity}}+1=3$, the scalar operator introduced by the ``fake primary effect'' dose not play a role. However, as soon as the curve reach $\Delta_{\rm x}=3$, the bound effectively indicates the scaling dimension of the second scalar operator in x-channel, which has a much higher scaling dimension. The bootstrap curve shows a jump. These fake jumps happens precisely at $\Delta_{\rm x}=3$. The jump is unphysical, and it will disappear by introducing a gap in the $\Delta^{l=1}_x$ channel, see Figure 9 of  \cite{Karateev:2019pvw}. If the jump corresponds to a physical CFT, the jump will survive as long as $\Delta^{l=1}_x$ gap is lower than the scaling dimension of the physical operator of the corresponding CFT, see     Figure 6 of \cite{Karateev:2019pvw}. 

\begin{figure}[h]
\includegraphics[width=12cm]{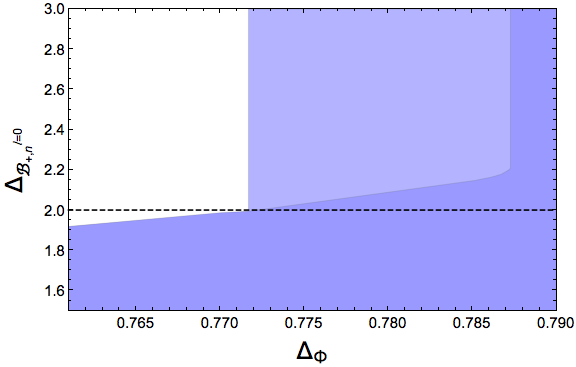}
\caption{Location of the jump as we vary the gap in the non-conserved current channel $\Delta_{\mathcal{F}^{j=1/2}_{+,n}}$. Lighter region corresponds to imposing $\Delta_{\mathcal{F}^{l=1}_{+,n}}\geq\Delta_{\text{unitarity}}$. The jump happens precisely at $\Delta_{\mathcal{B}^{l=0}_{+,n}}=2$, due to the ``fake primary'' effect. Darker region corresponds to $\Delta_{\mathcal{F}^{j=1/2}_{+,n}}\geq \Delta_{\text{unitarity}}+1.2$. }\label{fake1}
\end{figure}

\begin{figure}[h]
\includegraphics[width=12cm]{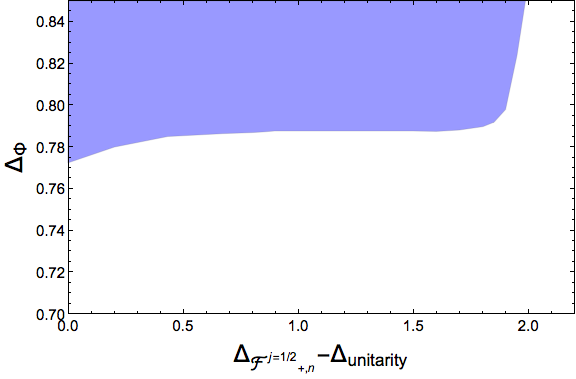}
\caption{Location of the jump as we vary the gap in the non-conserved current channel $\Delta_{\mathcal{F}^{j=1/2}_{+,n}}$. We have chose $\Delta_{\mathcal{B}^{l=0}_{+,n}}\geq 3$. }\label{fake2}
\end{figure}

In our $\cal N$=1 bootstrap program, a similar ``fake primary effect'' happens. The numerical conformal block for $\mathcal{F}^{j=1/2}_{+}$ multiplet, when $\Delta \rightarrow \Delta_{\text{unitarity}}$, looks like the conformal block for $\mathcal{F}^{j=1/2}_{+,\Delta_{\text{unitarity}}}$  plus a block for $\mathcal{B}^{l=0}_{+,\Delta=2}$. When we study $\cal N$=1 SCFT with SU(3) symmetry (and for the other member of the $F_4$ family), the bootstrap curve shows a jump at $\Delta_{\mathcal{B}^{l=0}_{+,n}}=2$, see Figure \ref{fake1}. This jump, after shifting a little bit, survives as we  impose $\Delta_{\mathcal{F}^{j=1/2}_{+,n}}\geq \Delta_{\text{unitarity}}+1.2$. This leads us to conjecture that such as jump corresponds to a physical CFT.  In Figure \ref{fake2}, we present how the location of the jump changes as we increase the $\Delta_{\mathcal{F}^{j=1/2}_{+,n}}$ gap.  

\section{Crossing equation for $\langle \phi^i\phi^j\phi^k\phi^l\rangle$, $\langle F^i F^j F^k F^l\rangle$ and $\langle \phi^i \phi^j F^k F^l\rangle$} \label{nonSUSYeqn}
As an intermediate step towards deriving the superconformal bootstrap equation involving four superfields $\langle\Phi^i\Phi^j\Phi^k\Phi^l\rangle$, we derived the mixed correlator bootstrap equation involving $\langle \phi^i\phi^j\phi^k\phi^l\rangle$, $\langle F^i F^j F^k F^l\rangle$ and $\langle \phi^i \phi^j F^k F^l\rangle$. For the moment, $\phi^i$ and $F^i$ are two different conformal primaries transforming in the same irreducible representation (labeled by $n$) of some flavor symmetry group $G$. Later, $\phi^i$ and $F^{i}$ would be identified as the superconformal primary and descendant of the supermultiplet $\Phi^i$ respectively. We also assume the theory preserves the time-reversal parity under which $\phi\rightarrow -
\phi$, while $F$ remains invariant. 

Let us first consider $\langle \phi^i\phi^j\phi^k\phi^l\rangle$. The correlator admits a conformal partial wave expansion as 

$$\langle \phi^i\phi^j\phi^k\phi^l\rangle=\frac{1}{x_{12}^{2\Delta_{\phi}}x_{34}^{2\Delta_{\phi}}}\sum_{I} P^{I}_{ijkl} \sum_{O\in I} (f_{\phi\phi I})^2  g^{\phi\phi;\phi\phi}_{\Delta_O,l_{O}}(u,v),$$
where $P^{I}_{ijkl}$ are so called projector which tells us how to decomposed the reducible representation $n\otimes n$ into irreducible representations. We also use the short-handed notation $f_{\phi\phi I}$ to denote $f_{\phi\phi O^{(I)}}$. Define a matrix $M$ by 
\be\label{MMM}
P^{R_1}_{kjil}=\sum_{R_2} M_{R_1 R_2} P^{R_2}_{ijkl}
\ee

Then the crossing equation are simply given by 
$$\sum_{R_2}P^{(R_2)} \bigg( \frac{1}{x_{12}^{2\Delta_{\phi}}x_{34}^{2\Delta_{\phi}}}\sum_{O\in R_2}(f_{\phi\phi R_2})^2g^{\phi\phi;\phi\phi}_{\Delta_O,l_{O}}(u,v)-\frac{1}{x_{23}^{2\Delta_{\phi}}x_{14}^{2\Delta_{\phi}}} \sum_{R_1} M_{R_1 R_2} \sum_{O\in R_1} (f_{\phi\phi R_1})^2   g^{\phi\phi;\phi\phi}_{\Delta_O,l_{O}}(v,u)\bigg)=0
$$
Define as usual the convolved conformal bock
$$
F^{ab;cd}_{\pm,\Delta,l}(u,v)=v^{\frac{\Delta_{c}+\Delta_{b}}{2}} g^{\Delta_{ab};\Delta_{cd}}_{\Delta,l}(u,v)\pm u^{\frac{\Delta_{c}+\Delta_{b}}{2}} g^{\Delta_{ab};\Delta_{cd}}_{\Delta,l}(v,u)
$$

The crossing equation becomes 

\be\label{crs1}
(\pm M^{T}-\mathbb{I}) \cdot \left(
\begin{array}{c}
 \sum_{O\in R_1}(f_{\phi\phi R_1})^2 F^{\phi\phi;\phi\phi}_{\pm \Delta_O,l_{O}}(u,v) \\
 \sum_{O\in R_2}(f_{\phi\phi R_2})^2 F^{\phi\phi;\phi\phi}_{\pm \Delta_O,l_{O}}(u,v) \\
 \dots \\
\end{array}
\right)=0.
\ee

Notice that not all there equation are independent. In practice, one should replace $(\pm M^{T}-\mathbb{I})$ by its row-reduced form (Mathematica function ``RowReduce'' can be used) to eliminate redundant equations. In the end, the number of independent equations from \eqref{crs1} equals to the number of irreps appearing in $n \otimes n$.

The crossing equations from $
\langle
\contraction{}{F_i(x_1)}{}{F_j(x_2)}
F_i(x_1)F_j(x_2)
\contraction{}{F_k(x_3)}{}{F_{l}(x_4)}
F_k(x_3)F_{l}(x_4)
\rangle=\langle
\contraction{}{F_k(x_3)}{}{F_j(x_2)}
F_k(x_3)F_j(x_2)
\contraction{}{F_i(x_1)}{}{F_{l}(x_4)}
F_i(x_1)F_{l}(x_4)
\rangle
\rangle.
$ can be similarly written as 
\be\label{crs2}
(\pm M^{T}-\mathbb{I}) \cdot \left(
\begin{array}{c}
 \sum_{O\in R_1}(f_{FF R_1})^2 F^{FF;FF}_{\pm \Delta_O,l_{O}}(u,v) \\
 \sum_{O\in R_2}(f_{FF R_2})^2 F^{FF;FF}_{\pm \Delta_O,l_{O}}(u,v) \\
 \dots \\
\end{array}
\right)=0.
\ee
Further more, the crossing equations from $
\langle
\contraction{}{\phi_i(x_1)}{}{F_j(x_2)}
\phi_i(x_1)F_j(x_2)
\contraction{}{\phi_k(x_3)}{}{F_{l}(x_4)}
\phi_k(x_3)F_{l}(x_4)
\rangle=\langle \contraction{}{\phi_k(x_3)}{}{F_j(x_2)}
\phi_k(x_3)F_j(x_2)
\contraction{}{\phi_i(x_1)}{}{F_{l}(x_4)}
\phi_i(x_1)F_{l}(x_4)
\rangle
$ are  
\be\label{crs3}
(\pm M^{T}-\mathbb{I}) \cdot \left(
\begin{array}{c}
 \sum_{O\in R_1^-}(f_{\phi F R_1^-})^2 F^{\phi F;\phi F}_{\pm \Delta_O,l_{O}}(u,v) \\
 \sum_{O\in R_2^-}(f_{\phi F R_2^-})^2 F^{\phi F; \phi F}_{\pm \Delta_O,l_{O}}(u,v) \\
 \dots \\
\end{array}
\right)=0.
\ee
Notice we have used $R_1^-$ to label the irreps here. The reason is that the operators appearing in $\phi \times F$ are parity odd under $\phi\rightarrow -
\phi$.  These operators are different from the parity even operators appearing in $\phi \times \phi$ or $F \times F$ OPE.

The remaining bootstrap equations come from the relation $
\langle
\contraction{}{F_i(x_1)}{}{F_j(x_2)}
F_i(x_1)F_j(x_2)
\contraction{}{\phi_k(x_3)}{}{\phi_{l}(x_4)}
\phi_k(x_3)\phi_{l}(x_4)
\rangle=\langle
\contraction{}{\phi_k(x_3)}{}{F_j(x_2)}
\phi_k(x_3)F_j(x_2)
\contraction{}{F_i(x_1)}{}{\phi_{l}(x_4)}
F_i(x_1)\phi_{l}(x_4)
\rangle
$. 
The LHS is 
\bea
\langle
\contraction{}{\phi_i(x_1)}{}{F_j(x_2)}
\phi_i(x_1)F_j(x_2)
\contraction{}{F_k(x_3)}{}{\phi_{l}(x_4)}
F_k(x_3)\phi_{l}(x_4)
\rangle= &&
\langle
\contraction{}{\phi_i(x_1)}{}{F_j(x_2)}
\phi_i(x_1)F_j(x_2)
\contraction{}{\phi_k(x_3)}{}{F_{l}(x_4)}
\phi_k(x_3) F_{l}(x_4)
\rangle \big |_{x_3\leftrightarrow x_4, k\leftrightarrow l}\nn\\
=&&(\ldots) \sum_{R_1} P^{R_1}_{ijlk}\big |_{k\leftrightarrow l} \sum_{O\in R_1}  (f_{\phi F R_1})^2 g^{\phi F; \phi F}_{\Delta_O,l_O}(u/v,1/v)\nn\\
=&&(\ldots) \sum_{R_1} P^{R_1}_{ijkl} (-1)^{b_{R_1}}\sum_{O\in R_1}  (-1)^{l} (f_{\phi F R_1})^2 g^{\phi F;  F \phi }_{\Delta_O,l_O}(u,v).
\nn\eea
As is explicit from the above derivation, $b_{R_1}$ is defined through the relation 
\be\label{plusminus}
P^{R_1}_{ijlk}= (-1)^{b_{R_1}}P^{R_1}_{ijkl} (-1)^{b_{R_1}}.
\ee
The RHS is 
\bea
\langle
\contraction{}{F_k(x_3)}{}{F_j(x_2)}
F_k(x_3)F_j(x_2)
\contraction{}{\phi_i(x_1)}{}{\phi_{l}(x_3)}
\phi_i(x_1)\phi_{l}(x_4)
\rangle=&&(\ldots) \sum_{R_2} P^{R_2}_{kjil} \sum_{O\in R_2} f_{FF R_2} f_{\phi\phi R_2}  g^{FF;\phi\phi}_{\Delta_O,l_{O}}(u,v)\nn\\=&&(\ldots) \sum_{R_2} \sum_{R_1} M_{R_2R_1}P^{R_1}_{ijkl} \sum_{O\in R_2} f_{FF R_2} f_{\phi\phi R_2}  g^{FF;\phi\phi}_{\Delta_O,l_{O}}(u,v).\nn
\eea
We have neglected the ($\ldots$) for simplicity. Similarly as \eqref{crs1}, we can write the crossing equation as 
\be\label{crs4}
 M^{T} \cdot \left(
\begin{array}{c}
 \sum_{O\in R_1} f_{FF R_1}f_{\phi\phi R_1} F^{FF;\phi\phi}_{\pm \Delta_O,l_{O}}(u,v) \\
 \sum_{O\in R_2} f_{FF R_2}f_{\phi\phi R_2} F^{FF;\phi\phi}_{\pm \Delta_O,l_{O}}(u,v)\\
 \dots \\
\end{array}
\right)\mp\left(
\begin{array}{c}
 \sum_{O\in R_1^-}(-1)^{b_{R_1}} (-1)^l (f_{\phi F R_1^-})^2 F^{\phi F;F \phi }_{\pm \Delta_O,l_{O}}(u,v) \\
 \sum_{O\in R_2^-}(-1)^{b_{R_2}}  (-1)^l (f_{\phi F R_2^-})^2 F^{\phi F; F \phi }_{\pm \Delta_O,l_{O}}(u,v) \\
 \dots \\
\end{array}
\right)=0.
\ee

Altogether, \eqref{crs1},  \eqref{crs2},  \eqref{crs3} and  \eqref{crs4} give us the full set of crossing equations. We can write them in the more familiar form which could be converted to a semi-definite programming problem.
Define
 $$K^{\pm}=\text{RowReduce}[\pm M^T-\mathbb{I}]$$
 and also define $\vec{m}_R$ and $\vec{k}^{\pm}_R$ to be the $R$-th column of $M$ and $K^{\pm}$ respectively.  Similarly, define $\vec{i}_R$ to be the $R$-th column of the identity martrix $\mathbb{I}$.

\be\label{crsall1}
V^{+}_{R}=\left(
\begin{array}{c}
\vec{k}^{+}_R \times \left(
\begin{array}{cc}
 F_{+}^{\phi\phi;\phi\phi} & 0 \\
 0 & 0 \\
\end{array}
\right)\\
\vec{k}^{-}_R \times \left(
\begin{array}{cc}
 F_{-}^{\phi\phi;\phi\phi} & 0 \\
 0 & 0 \\
\end{array}
\right) \\
\vec{k}^{+}_R \times \left(
\begin{array}{cc}
0 & 0 \\
 0 &  F_{+}^{FF;FF}  \\
\end{array}
\right)\\
\vec{k}^{-}_R \times \left(
\begin{array}{cc}
 0 & 0 \\
 0 &  F_{+}^{FF;FF}  \\
\end{array}
\right) 
\\
 \left(
\begin{array}{cc}
 0 & 0 \\
 0 & 0 \\
\end{array}
\right) \\
 \left(
\begin{array}{cc}
 0 & 0 \\
 0 & 0 \\
\end{array}
\right)\\
 \vec{m}^{}_R \times \left(
\begin{array}{cc}
 0 & \frac{1}{2} F_{+}^{FF;\phi\phi} \\
 \frac{1}{2} F_{+}^{FF;\phi\phi} &  0  \\
\end{array}
\right) \\
 \vec{m}^{}_R \times \left(
\begin{array}{cc}
 0 & \frac{1}{2} F_{-}^{FF;\phi\phi} \\
 \frac{1}{2} F_{-}^{FF;\phi\phi} &  0  \\
\end{array}
\right) 
\\
\end{array}
\right),
 \ee
and 
\be\label{crsall2}
V^{-}_{R}=\left(
\begin{array}{c}
 \vec{0} \\
 \vec{0} \\
 \vec{0} \\
 \vec{0} \\
 \vec{k}^{+}_R \times F_{+}^{\phi F;\phi F} \\
  \vec{k}^{-}_R \times F_{-}^{\phi F;\phi F}  \\
  \vec{i}^{+}_R \times (-1)(-1)^{b_R} (-1)^l  F_{+}^{ \phi F; F \phi} \\
 \vec{i}^{+}_R \times  (-1)^{b_R} (-1)^l F_{+}^{ \phi F; F \phi} 
\end{array}
\right)
\ee 
It may seems from the above expression that we have in total $8p$  crossing equation, suppose $p$ is the number of irreps in $n\otimes n$. This is however not true. As mentioned already,  $K^{\pm}=\text{RowReduce}[\pm M^T-\mathbb{I}]$ have empty. We should delete these rows before plugging in \eqref{crsall1} and   \eqref{crsall2}. In fact, $K^+$ and $K^{-}$ together have $p$ non-zero rows. We have $5p$ lines of crossing equations. 

Notice also that the operators appearing in $\phi^i\times\phi^j$ and $F^i\times F^j$ are selected according to the signature $b_R$.  When $b_R=+/-$, only even/odd spin operators are allowed. We need to take this into account when using the conformal block $V^{+}_{R}$.

\section{SUSY crossing equation for $\langle \Phi^i \Phi^j \Phi^k \Phi^l\rangle$}\label{susyeqn}

When the theory have supersymmetry, the operator product expansion (OPE) coefficients in $\phi \times \phi$, $F \times F$ and $\phi \times F$ are related to each other. Since flavor symmetry commutes with the superconformal symmetry, we could use the same OPE relations as in \cite{Rong:2018okz}. Plugging them into the non-SUSY crossing equation, we obtain the following corssing equantions.  
For $\mathcal{B}_+$ multiplets in representation $R$,
\be\label{SUSYcrsall1}
V^{\mathcal{B}_+}_{R}=\left(
\begin{array}{c}
\vec{k}^{+}_R \times F_{+,\Delta,l}^{\phi\phi;\phi\phi} \\
\vec{k}^{-}_R \times F_{-,\Delta,l}^{\phi\phi;\phi\phi} \\
\vec{k}^{+}_R \times c_2 F_{+,\Delta+1,l}^{\phi F;\phi F}  \\
\vec{k}^{-}_R \times c_2 F_{-,\Delta+1,l}^{\phi F;\phi F}  \\
 \vec{m}_R \times c_1 F_{+,\Delta,l}^{F F; \phi \phi} -  \vec{i}_R \times c_2 (-1)^{b_R} (-1)^l F_{+,\Delta+1,l}^{\phi F; F \phi} \\
   \vec{m}_R \times c_1 F_{-,\Delta,l}^{F F; \phi \phi} +  \vec{i}_R \times c_2 (-1)^{b_R} (-1)^l  F_{-,\Delta+1,l}^{\phi F; F \phi}   
\end{array}
\right)_{\Delta_F=\Delta_\phi+1},
\ee 
with 
\be
c_1=\frac{\left(2 \Delta_\phi-\Delta -l-1\right) \left(2 \Delta_\phi-\Delta +l\right)}{2 \Delta_\phi \left(2 \Delta_\phi-1\right)}, \quad\quad\quad\quad \text{ }   c_2=\frac{(\Delta -1) (\Delta -l-1) (\Delta +l)}{4 (2 \Delta -1) \Delta_\phi \left(2 \Delta_\phi-1\right)}.\nn
\ee
There is a spin selection rule similar to the non-SUSY case, when $b_R=+/-$,  only $\mathcal{B}^{l}_+$ multiplets with even/odd $l$ are allowed.
 
For $\mathcal{B}_-$ multiplets in representation $R$,
\be\label{SUSYcrsall2}
V^{\mathcal{B}_-}_{R}=\left(
\begin{array}{c}
\vec{k}^{+}_R \times F_{+,\Delta+1,l}^{\phi\phi;\phi\phi} \\
\vec{k}^{-}_R \times F_{-,\Delta+1,l}^{\phi\phi;\phi\phi} \\
\vec{k}^{+}_R \times d_2 F_{+,\Delta,l}^{\phi F;\phi F}  \\
\vec{k}^{-}_R \times d_2 F_{-,\Delta,l}^{\phi F;\phi F}  \\
 \vec{m}_R \times d_1 F_{+,\Delta+1,l}^{F F; \phi \phi} -  \vec{i}_R \times d_2 (-1)^{b_R} (-1)^l  F_{+,\Delta,l}^{\phi F; F \phi} \\
   \vec{m}_R \times d_1 F_{-,\Delta+1,l}^{F F;\phi \phi} +  \vec{i}_R \times d_2 (-1)^{b_R} (-1)^l  F_{-,\Delta,l}^{\phi F; F \phi}   
\end{array}
\right)_{\Delta_F=\Delta_\phi+1},
\ee 
with
\be
 d_1=\frac{\left(2 \Delta_\phi+\Delta -l-3\right) \left(2 \Delta_\phi+\Delta +l-2\right)}{2 \Delta_\phi \left(2 \Delta_\phi-1\right)},\quad\quad \text{ } \text{ } d_2=\frac{(2 \Delta -1) (\Delta -l-1) (\Delta +l)}{(\Delta -1) \Delta_\phi \left(2 \Delta_\phi-1\right)},\nn
\ee
There is a spin selection rule similar to the non-SUSY case, when $b_R=+/-$, only  $\mathcal{B}^{l}_-$ multiplets with even/odd $l$ are allowed.

For $\mathcal{F}_+$ multiplets in representation $R$,

\be\label{SUSYcrsall3}
V^{\mathcal{F}_+}_{R}=\left(
\begin{array}{c}
\vec{k}^{+}_R \times F_{+,\Delta_l,l+1}^{\phi\phi;\phi\phi} \\
\vec{k}^{-}_R \times F_{-,\Delta_l,l+1}^{\phi\phi;\phi\phi} \\
\vec{k}^{+}_R \times f_2 F_{+,\Delta_l,l}^{\phi F;\phi F}  \\
\vec{k}^{-}_R \times f_2 F_{-,\Delta_l,l}^{\phi F;\phi F}  \\
 \vec{m}_R \times f_1 F_{+,\Delta_l,l+1}^{F F; \phi \phi} -  \vec{i}_R \times f_2 (-1)^{b_R} (-1)^l  F_{+,\Delta_l,l}^{\phi F; F \phi} \\
   \vec{m}_R \times f_1 F_{-,\Delta_l,l+1}^{F F; \phi \phi} +  \vec{i}_R \times f_2 (-1)^{b_R} (-1)^l  F_{-,\Delta_l,l}^{\phi F; F \phi}   
\end{array}
\right)_{\Delta_F=\Delta_\phi+1}
\ee 
with
\be
f_1=\frac{\left(-2 \Delta_\phi-\Delta_l +l+4\right) \left(-2 \Delta_\phi+\Delta_l +l+1\right)}{2 \Delta_\phi \left(2 \Delta_\phi-1\right)},\quad f_2=\frac{(2 l+1) (\Delta_l -l-2) (\Delta_l +l)}{2 (l+1) \Delta_\phi \left(2 \Delta_\phi-1\right)}.\nn
\ee

There is a spin selection rule similar to the non-SUSY case, when $b_R=+/-$, only  $\mathcal{F}^{j}_+$ multiplets with even/odd $j-1/2$ are allowed.

\be\label{SUSYcrsall4}
V^{\mathcal{F}_-}_{R}=\left(
\begin{array}{c}
\vec{k}^{+}_R \times F_{+,\Delta_l,l}^{\phi\phi;\phi\phi} \\
\vec{k}^{-}_R \times F_{-,\Delta_l,l}^{\phi\phi;\phi\phi} \\
\vec{k}^{+}_R \times e_2 F_{+,\Delta_l,l+1}^{\phi F;\phi F}  \\
\vec{k}^{-}_R \times e_2 F_{-,\Delta_l,l+1}^{\phi F;\phi F}  \\
 \vec{m}_R \times e_1 F_{+,\Delta_l,l}^{F F; \phi \phi} -  \vec{i}_R \times e_2 (-1)^{b_R} (-1)^l  F_{+,\Delta_l,l+1}^{\phi F; F \phi} \\
   \vec{m}_R \times e_1 F_{-,\Delta_l,l}^{F F; \phi \phi} +  \vec{i}_R \times e_2 (-1)^{b_R} (-1)^l  F_{-,\Delta_l,l+1}^{\phi F; F \phi}   
\end{array}
\right)_{\Delta_F=\Delta_\phi+1},
\ee 

For $\mathcal{F}_-$ multiplets in representation $R$,

\be
e_1=\frac{\left(2 \Delta_\phi-\Delta_l +l+1\right) \left(2 \Delta_\phi+\Delta_l +l-2\right)}{2 \Delta_\phi \left(2 \Delta_\phi-1\right)}, \quad\quad\text{ } \text{ } e_2=\frac{(l+1) (\Delta_l -l-2) (\Delta_l +l)}{2 (2 l+1) \Delta_\phi \left(2 \Delta_\phi-1\right)}.\nn
\ee
There is a spin selection rule similar to the non-SUSY case, When $b_R=+/-$, only  $\mathcal{F}^{j}_-$ multiplets with odd/even $j-1/2$ are allowed.

The coefficients $\{c_1,c_2,d_1,d_2,e_1,e_2,f_1,f_2\}$ are the same coefficients used derived in \cite{Rong:2018okz} by analyzing the allowed 3-pt structures of $\cal N$=1 SCFT \cite{Park:1999cw}.

As we have seen, the bootstrap equation depends on the matrix 
\be
 M_{R_1 R_2}
\ee
defined in \eqref{MMM}, which can be calculated explicitly using projectors. 

The projectors for multiplying two n dimensional irreps of  $S_N$ group with N$>$3 could be found in \cite{Rong:2017cow}, from which we find 
\be
M=\left(
\begin{array}{cccc}
 \frac{1}{N+1} & \frac{1}{N+1} & \frac{1}{N+1} & -\frac{1}{N+1} \\
 1 & -\frac{1-N}{N} & -\frac{1}{N} & \frac{1}{N} \\
 -\frac{-(N+1)^2+N+3}{2 (N+1)} & -\frac{(N+1)^2-N-3}{2 N (N+1)} & -\frac{-(N+1)^2+N-1}{2 N (N+1)} & -\frac{-(N+1)^2+N+3}{2 N (N+1)} \\
 -\frac{N}{2} & \frac{1}{2} & \frac{1}{2} & \frac{1}{2} \\
\end{array}
\right).
\ee
The representations are ordered as $\{S,n,T',A\}$.  The signature $b_R$ define in \eqref{plusminus} are $\{+,+,+,-\}$ respectively. The external superfield $\Phi$ transforms in the $n=N-1$ dimensional representation of $S_N$. ``$S$'' is the singlet, ``$A$'' is the antisymmetric representation. Since $S_N$ group is a subgroup of $SO(n)$, the symmetric traceless representation ``T'' of $SO(n)$ branches into ``$n$'' and ``$T'$''.

When we consider the group $S_3$, the irrep $T'$ vanish, and the M matrix become a $3\times3$, which equals the M matrix for O(2) \cite{Rong:2017cow}:
\be
M=\left(
\begin{array}{ccc}
 \frac{1}{N} & \frac{1}{2} & -\frac{1}{2} \\
 -\frac{-N^2-N+2}{N^2} & -\frac{2-N}{2 N} & -\frac{-N-2}{2 N} \\
 -\frac{N-1}{N} & \frac{1}{2} & \frac{1}{2} \\
\end{array}\right)_{N=2}=\left(
\begin{array}{ccc}
 \frac{1}{2} & \frac{1}{2} & -\frac{1}{2} \\
 1 & 0 & 1 \\
 -\frac{1}{2} & \frac{1}{2} & \frac{1}{2} \\
\end{array}
\right).
\ee
 
The projectors for multiplying two adjoint irreps of $SU(N)$ group with N$>$3 (the group SU(3) belongs to the $F_4$ family, and will be discussed later) could be found in Table 9.4 of \cite{Cvitanovic:2008zz}, from which we obtain the matrix M to be 
\be
M=\left(
\begin{array}{cccccc}
 -\frac{1}{1-N^2} & -\frac{1}{1-N^2} & -\frac{1}{1-N^2} & -\frac{1}{1-N^2} & -\frac{1}{N^2-1} & -\frac{1}{N^2-1} \\
 1 & -\frac{12-N^2}{2 \left(N^2-4\right)} & -\frac{1}{N-2} & \frac{1}{N+2} & -\frac{1}{2} & \frac{2}{(N-2) (N+2)} \\
 -\frac{3 N^2-N^3}{4 (N-1)} & -\frac{N^3-3 N^2}{4 \left(N^2-3 N+2\right)} & -\frac{-N^2+N-2}{4 \left(N^2-3 N+2\right)} & -\frac{3-N}{4 (N-1)} & \frac{3 N-N^2}{4 (N-1)} & \frac{N^2-3 N}{4 (N-2) (N-1)} \\
 -\frac{-N^3-3 N^2}{4 (N+1)} & -\frac{-N^3-3 N^2}{4 \left(N^2+3 N+2\right)} & -\frac{-N-3}{4 (N+1)} & -\frac{-N^2-N-2}{4 \left(N^2+3 N+2\right)} & \frac{N^2+3 N}{4 (N+1)} & \frac{N^2+3 N}{4 (N+1) (N+2)} \\
 -1 & -\frac{1}{2} & -\frac{1}{N} & \frac{1}{N} & \frac{1}{2} & 0 \\
 \frac{1}{2} \left(4-N^2\right) & 1 & \frac{N+2}{2 N} & \frac{N-2}{2 N} & 0 & \frac{1}{2} \\
\end{array}
\right).
\ee
The representations are ordered as $\{S,n=Adj_s,T_1,T_2, Adj_a, A'\}$. The signature $b_R$ define in \eqref{plusminus} are $\{+,+,+,+,-,-\}$ respectively. Let us take SU(5) as an example to understand what these irreps are. The product of two adjoint irreps are (in terms of Dynkin label)
\bea
(1,0,0,1)\times (1,0,0,1)&\rightarrow& (0,0,0,0)_S+(1,0,0,1)_{Adj_s}+(0,1,1,0)_{T_1}+(2,0,0,2)_{T_2}\nonumber\\
&&+(1,0,0,1)_{Adj_a}+[(0,1,0,2)+(2,0,1,0)]_A
\eea
Notice the two irreps (0,1,0,2) and (2,0,1,0) are conjugate of each other. When considering external operator to be real superfields, we should treat them as a single representation.

The projectors for multiplying two irreps ``n'' of $F_4$ family of Lie groups could be found in \cite{Pang:2016xno}. The details of the construction of these projectors could be found in \cite{Cvitanovic:2008zz}. Using these projectors we find that 
\be
M=\left(
\begin{array}{ccccc}
 \frac{1}{n} & \frac{1}{n} & \frac{1}{n} & -\frac{1}{n} & -\frac{1}{n} \\
 1 & \frac{2-n}{2 n+4} & \frac{2}{n+2} & -\frac{1}{2} & \frac{4}{n+2} \\
 \frac{1}{2} \left(n-1-\frac{2}{n}\right) & \frac{(n-2) (n+1)}{n (n+2)} & \frac{n^2-4 n-4}{2 n (n+2)} & 1+\frac{1}{n} & \frac{(n-2)^2}{2 n (n+2)} \\
 \frac{6-3 n}{n+10} & \frac{6-3 n}{2 n+20} & \frac{6}{n+10} & \frac{1}{2} & 0 \\
 -\frac{n^2+3 n+2}{2 n+20} & \frac{2 (n+1)}{n+10} & \frac{n-2}{2 (n+10)} & 0 & \frac{1}{2} \\
\end{array}
\right).
\ee
The integer ``n'' could take four different values, as shown in Table \ref{F4table}. The representations appear in $n\times n$ could be labeled by $\{S,n,T',Adj, A'\}$. This is the order of irreps that we used when defining M above. For the dimensions of these irreps, check \cite{Pang:2016xno}. The signature $b_R$ defined in \eqref{plusminus} are $\{+,+,+,-,-\}$ respectively.

\end{document}